\documentclass[reprint]{natureprintstyle}

\usepackage{amsmath}
\usepackage{graphicx,amssymb}
\usepackage{comment}
\usepackage{color}
\title{Evidence of The Anomalous Fluctuating Magnetic State by Pressure Driven 4f Valence Change in EuNiGe$_3$}

\author{{K.~Chen$^{1,*}$, C.~Luo$^{2}$, Y.~Zhao$^3$ F.~Baudelet$^{4}$, A.~Maurya$^{5}$, A.~Thamizhavel$^5$, U.K.~R\"{o}\ss{}ler$^6$, D.~Makarov$^7$, and F.~Radu$^{2,*}$}}

\begin{document}
\maketitle

\begin{affiliations}
 \item National Synchrotron Radiation Laboratory, University of Science and Technology of China, Hefei 230026, Anhui, China
 \item Helmholtz-Zentrum Berlin f\"{u}r Materialien und Energie, Albert-Einstein-Strasse 15, 12489, Berlin, Germany
\item Center for High Pressure Science and Technology Advanced Research (HPSTAR), Shanghai 201203, China
 \item Synchrotron SOLEIL, L'Orme des Merisiers, Saint-Aubin-BP48, 91192 GIF-sur-Yvette Cedex, France
 \item Department of Condensed Matter Physics and Materials Science, Tata Institute of Fundamental Research, Colaba, Mumbai 400005, India\\
 \item Leibniz-Institut für Festk\"{o}rper- und Werkstoffforschung Dresden e. V. (IFW Dresden), 01069 Dresden, Germany
 \item Helmholtz-Zentrum Dresden-Rossendorf e.V., Institute of Ion Beam Physics and Materials Research, 01328 Dresden, Germany
\end{affiliations}

\begin{abstract}
\large\textbf{ABSTRACT}
\\
In rare-earth compounds with valence fluctuation, the proximity of the 4f level to the Fermi energy leads to instabilities of the charge configuration and the magnetic moment. Here, we provide direct experimental evidence for an induced magnetic polarization of the Eu$^{3+}$ atomic shell with J=0, due to intra-atomic exchange and spin-orbital coupling interactions with Eu$^{2+}$ atomic shell. By applying external pressure, a transition from antiferromagnetic to a fluctuating behavior in a EuNiGe$_3$ single crystals is probed. Magnetic polarization is observed for both valence states of Eu$^{2+}$ and Eu$^{3+}$ across the entire pressure range. The anomalous magnetism is discussed in terms of a homogeneous intermediate valence state where frustrated Dzyaloshinskii-Moriya couplings are enhanced by the onset of spin-orbital interaction and engender a chiral spin-liquid-like precursor.
\end{abstract}
\newpage
\begin{comment}
\begin{center}
\large\textbf{TOC Graphic}
\end{center}
\begin{figure}[h]
\centering
\includegraphics[width=0.6\linewidth]{TOC.pdf}
%\caption{\small{  } }
 \label{fig:toc1}
\end{figure}

\newpage
\end{comment}

\section*{Introduction}\hspace*{\fill}\\
\\
Solid state systems can undergo electronic transitions leading to intermediate or mixed valencies creating systems of ions with co-existing and, thus, correlated electronic configurations~\cite{Riseborough2016}. Looking beyond the single ion, the understanding of collective phenomena, like magnetic ordering or superconductivity, in such strongly correlated electronic systems remains a major problem in condensed matter physics~\cite{Khomskii1979}. If the two co-existing limiting configurations own qualitatively different magnetic states, a long-range ordered magnetic ground state may disappear or be replaced by a hidden or exotic magnetic order. Europium compounds with valence-fluctuating states provide a fruitful realization of such a valence transition. The divalent Eu$^{2+}$ state with 4f$^7$ (L=0, S=7/2, and J=7/2) has a large pure spin-moment, while the trivalent Eu$^{3+}$ with 4f$^6$ configuration (L=3, S=3, and J=0) is magnetically invisible. As the energy difference between Eu$^{2+}$ and Eu$^{3+}$ valence is not large~\cite{Bauminger1973}, orbital intermixing can be achieved by applying external pressure or chemical substitution~\cite{Loula2012,Paramanik2016, Onuki2017}.  Thus, co-existence of electronic configurations with energy differences in the thermal range can be achieved. By increase of the trivalent Eu at the expense of the divalent Eu, transitions from magnetically ordered to the paramagnetic state are expected, like in Ce and Yb-based materials~\cite{Gegenwart2008}.  However, in the transition region the intermediate valency of the magnetic sites and a complex character of the intersite couplings may create novel magnetic behavior, as both are based on a strongly correlated electronic structure~\cite{Souza-Neto2009,Uchima2014,Loula2012}.\\

The 4f$^6$ configuration owns a spin-polarization with an identical but oppositely aligned orbital moment, and in addition the $J > 0$ excitation of the 4f$^6$-shell give rise to Van Vleck (para)-magnetism \cite{Vleck1932}. For the collective behavior, it has been theoretically suggested that such Van Vleck ions can contribute with a particular (anisotropic) intersite magnetic exchange \cite{HuangVanVleck1969,Palermo1980}, which could drive a hidden spin-ordering \cite{Johannes2005}. Up to now, observation of hidden magnetic correlations between individual ions with 4f$^6$ or also 5f$^6$ configurations is rare~\cite{Magnani2015, Binh2013, Skaugen2015} and considered to originate from intersite exchange coupling mechanisms and the presence of a spin-polarized matrix acting on the Van Vleck ions\cite{Magnani2015,Skaugen2015,Binh2013,Takikawa2010, Ruck2011}. The admixture of the trivalent Eu also implies a modified orbital structure which could by -atomic exchange and spin-orbital interactions affect the magnetic ordering. In this work, we investigate a magnetic europium compound with non-centrosymmetric lattice structure, which allows for the presence of the antisymmetric Dyzaloshinskii-Moriya interactions (DMIs)~\cite{Dzyaloshinskii1958,Moriya1960}. Magnetically, the EuNiGe$_3$ exhibits a complex magnetic behavior below the Neel temperature (measured to be 13.2 K, see Supplementary information), including a incommensurate helicoidal  magnetic structure at 3.6 K \cite{Ryan2016}. These couplings cause effective chiral couplings~\cite{Dzyaloshinskii1964} that frustrate homogeneous magnetic states and preclude conventional ordering according to the fundamental Landau theory of phase transitions~\cite{ToledanoToledano1987}. Instead an intermediate chiral liquid-like or partially ordered state may appear~\cite{Roessler2006,Wilhelm2011}, as experimentally found in chiral helimagnets like MnSi and FeGe under pressure~\cite{Pfleiderer2004,Barla2015}.\\

We report on a pressure induced electronic phase transition in an antiferromagnetic and metallic compound, EuNiGe$_3$~\cite{Goetsch2013,Maurya2014,Maurya2015,Fabreges2016,Ryan2016} where a change of valence from dominating Eu$^{2+}$ to an intermediate valence close to  Eu$^{2.5+}$ causes the appearance of a fluctuating magnetic state. This state is anomalous as it displays no magnetically homogeneous long-range order (LRO), but is not paramagnetic either. Its thermal fluctuations can be characterized and quantified with the model for superparamagnetic (SPM) behavior. We argue that the observation of this unusual magnetism is evidence for a strongly-correlated electronic system with partial magnetic order under the influence of chiral magnetic coupling caused by spin-orbit interactions. The tools used to detect the valence transition and the evolution of the fluctuating state are temperature- and pressure-dependent x-ray absorption spectroscopy (XAS) and x-ray magnetic circular dichroism (XMCD) at Eu L$_2$-edge, which are able to distinguish the polarization of the 5d orbital channels. Complementary to the Van Vleck paramagnetism characteristic for materials containing mainly Eu$^{3+}$ ions, the polarization of 5d channels of Eu$^{3+}$ states mirrors the magnetic behavior of Eu$^{2+}$ under pressure, showing the same transition from AFM to a SPM-like behavior at about 30 GPa. In addition we observe a clear electronic phase transition of the Eu$^{3+}$ as evidenced by a sudden linewith change at the critical pressure.  Our results provide a direct evidence of intra-atomic exchange and spin-orbital interactions between the 5d channels of Eu$^{2+}$ and Eu$^{3+}$ contributions, which are essential to be considered when interpreting the physical properties of strongly correlated electronic systems.

\section*{Results}
\begin{figure}[t]
\centering
\includegraphics[width=0.8\linewidth]{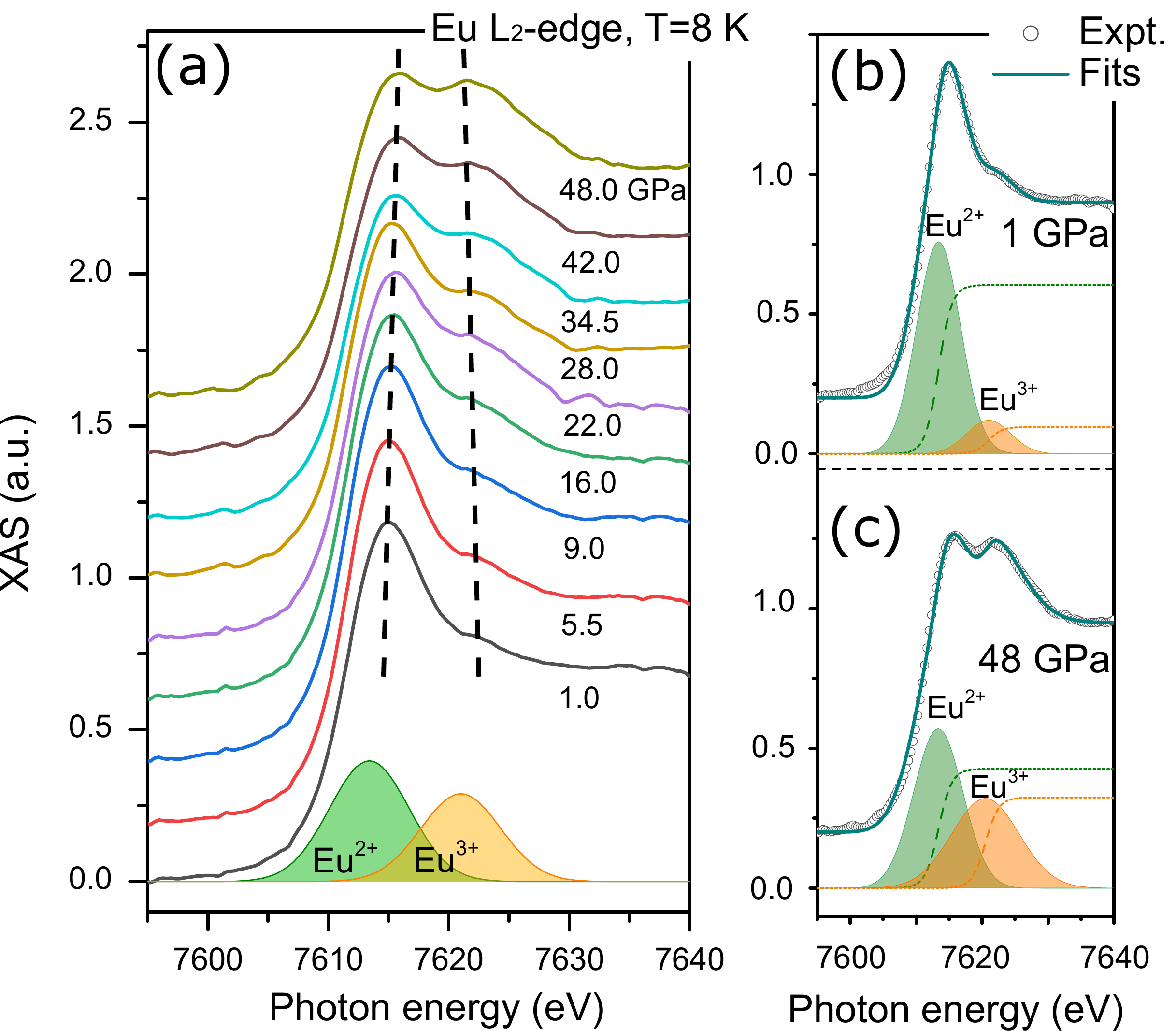}
\caption{\small{(a)~The XAS of Eu L$_{2}$-edge at 8~K under pressure up to 48 GPa, and the spectra of P=1~GPa (b) and 48~GPa (c) fitted with the combination of the spectra of $\rm Eu^{2+}$ and $\rm Eu^{3+}$ with Gaussian-type lineshapes (thin solid lines). The dashed curves represent the integral background. 
 } }
 \label{fig:1}
\end{figure}

\hspace*{\fill}\\
\\
The XAS spectra at Eu L$_2$-edge were taken at T=8~K for a pressure range up to 48~GPa, as shown in Fig.~\ref{fig:1}a. The quadrupolar (2p-4f) contributions which generally appear at the pre-edge~\cite{Souza-Neto2016} were not observed suggesting that the spectra are dominated by the dipolar contributions (2p$_{1/2}$-5d$_{3/2}$). The two contributions, shifted by $\rm \sim 7.7~eV$ against each other (dashed lines in Fig.~\ref{fig:1}a), belong to the 5d orbital channels of Eu$^{2+}$ (4f$^7$) and Eu$^{3+}$ (4f$^6$) states, respectively. This result, showing the coexistence of Eu$^{2+}$ and Eu$^{3+}$ levels, indicates that the valence fluctuation in EuNiGe$_3$ takes place. A decrease of the Eu$^{2+}$ content with a concomitant increase of the Eu$^{3+}$ contribution is observed, evidencing the valence increasing under pressure. This valence change can be anticipated to increase as a function of pressure since electrons will be transferred out of the 4f shells into the conduction band~\cite{debessai2009}.\\

To evaluate the mean values of Eu valence, the spectra are analyzed by assigning Gaussian lineshapes to the Eu$^{2+}$ and Eu$^{3+}$ contributions, each with a \textit{tanh}-type background, as shown in Fig.~\ref{fig:1}b-c for the pressure of 1 and 48 GPa, respectively. The weighted sum, 86\% Eu$^{2+}$ and 14\% Eu$^{3+}$ for 1 GPa and 57\% Eu$^{2+}$ and 43\% Eu$^{3+}$ for 48 GPa, of the simulated curves describes the EuNiGe$_3$ spectrum very well. The Eu mean valence can be derived from: $\rm \nu=2+I(Eu^{3+})/[I(Eu^{3+}) + I (Eu^{2+})]$, where $\rm I(Eu^{3+}) $ and $\rm I(Eu^{2+})$ denote the integrated intensities of the Eu$^{3+}$ and Eu$^{2+}$ components. Applying the fitting procedure, the Eu mean valence $\nu$ as a function of pressure have been extracted and shown in Fig.~\ref{fig:3}a. The results suggest that the Eu ion has a lower mean valence of $\nu$=2.13(3) at T=8 K and 1 GPa and a much higher value of $\nu$=2.43(3) when the pressure increases to 48 GPa. The value at low pressure is in good agreement with the literature report $\nu=$2.09~\cite{Utsumi2018}. An substantial enhancement of $\nu$ can be seen up to 48~GPa, except for  P$<$10~GPa below which a nearly constant value of $\sim$2.13(3) is preserved. This is in agreement with previous results showing no valence change up to 8~GPa according to the electrical resistivity measuremts of EuNiGe$_3$~\cite{Uchima2014}. \\

\begin{figure}[t]
\centering
\includegraphics[width=.8\linewidth]{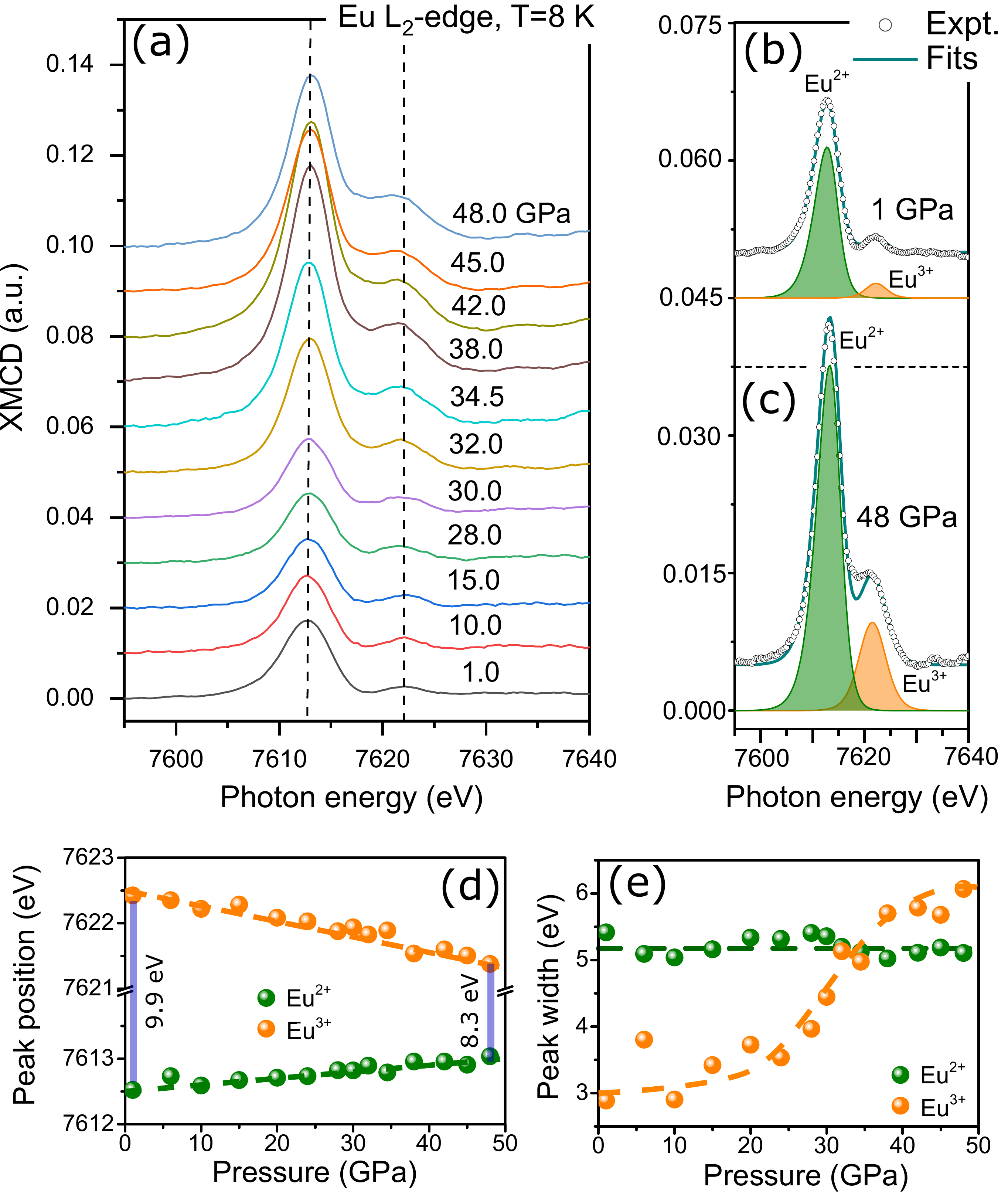}
\caption{\small{(a)~Pressure-dependent Eu L$_2$-edge XMCD spectra of bulk EuNiGe$_3$ up to 48 GPa at T=8~K and $\mu_0H$=1.4~T, normalized to the XAS intensity, and the XMCD spectra at P=1 (b) and 48 GPa (c) fits with the combination of the spectra of $\rm Eu^{2+}$ and $\rm Eu^{3+}$ with asymmetric double sigmoidal-type lineshapes (d) the resonances peak position as a function of pressure, showing a pressure-induced compression effect  (e) the FWHM for the Eu$^{2+}$ and Eu$^{3+}$. At the critical pressure the FWHM of Eu$^{3+}$ resonance exhibit a significant change, demonstrating the occurrence of a pressure-induced electronic phase transition. }}
 \label{fig:2}
\end{figure}

\begin{figure}[t]
\centering
\includegraphics[width=0.8\linewidth]{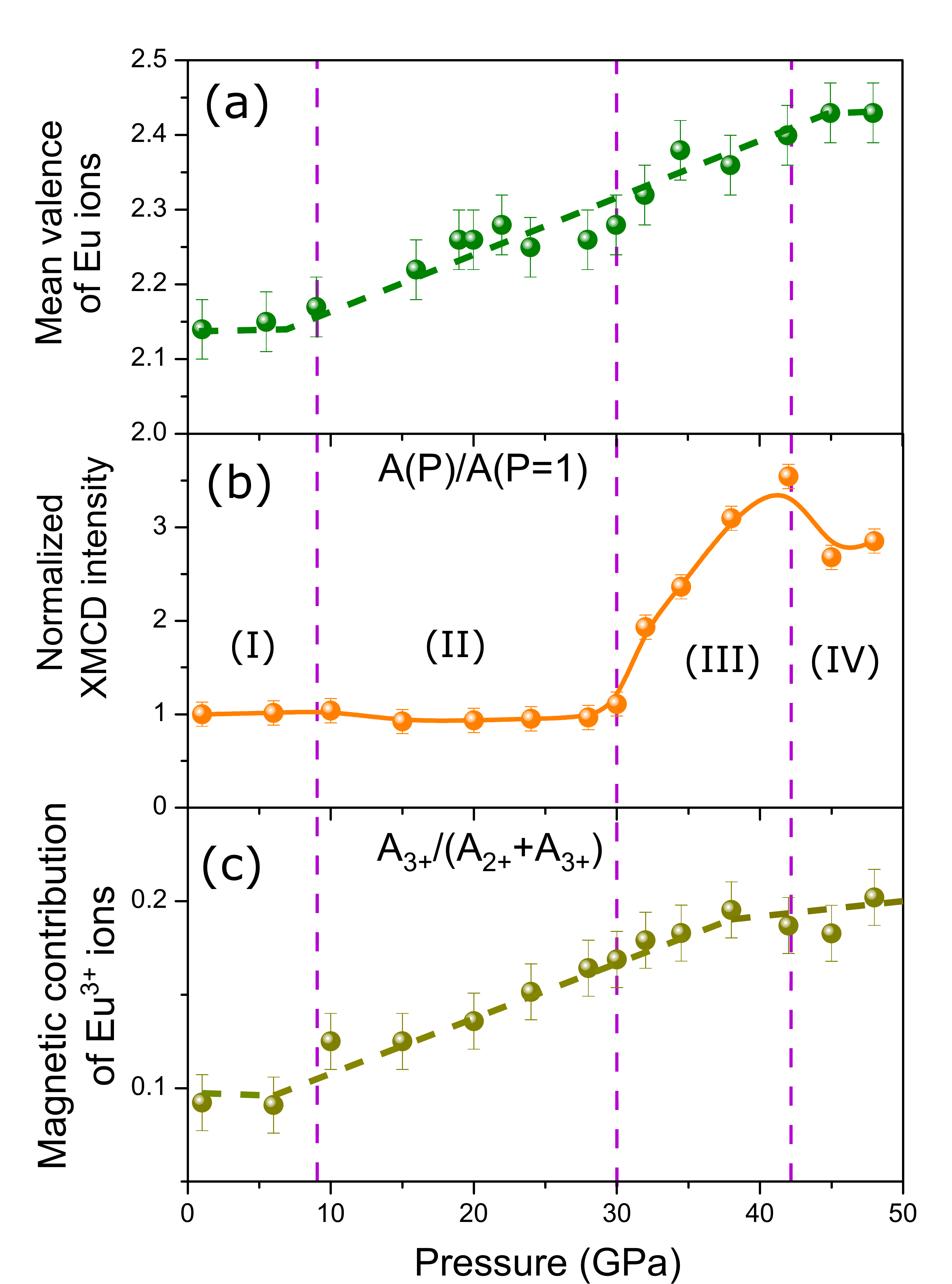}
\caption{\small{The Eu mean valence $\nu$ (a), the normalized XMCD intensity (b) with a non-uniform behavior, and the relative magnetic contribution from 5d channel of Eu$^{3+}$, $A_{3+}/(A_{2+}+A_{3+})$ (c), obtained as a function of the pressure at $\mu_0H$=1.4~T and T=8~K up to 48~GPa.} }
 \label{fig:3}
\end{figure}

The L$_2$-edge XMCD spectra from the dipolar transition (2p$^6$5d$^0$ $\rightarrow$ 2p$^5$5d$^1$) reflect the polarization of the 5d empty-state orbitals in the conduction band. The pressure dependent Eu L$_2$-edge XMCD spectra and their lineshape analysis(FWHM and energy position), which were recorded at T=8~K and $\mu_0H$=1.4~T and normalized to the XAS intensity, are presented in Fig.~\ref{fig:2}a. Similar to the XAS, two well defined peaks in the L$_{2}$ transitions from Eu$^{2+}$ and Eu$^{3+}$ channels are clearly present in the XMCD spectra and are drastically affected by pressure. This undoubtedly indicates that the Eu 5d orbital are magnetically polarized in both Eu$^{2+}$ and Eu$^{3+}$ channels. The magnetic contribution from both channels can be well separated as shown in Fig.~\ref{fig:2}b-c for pressure of 1 and 48~Gpa, respectively. The area of the two peaks from Eu$^{2+}$ and Eu$^{3+}$ electronic states are denoted as A$_{2+}$ and A$_{3+}$, respectively, to further investigate the pressure dependence of the magnetic polarization from different 5d orbital channels. The peak positions of the XMCD spectra are slightly below the XANES peaks, similar to  other Eu- and Sm- based fluctuating-valence materials of EuN~\cite{Binh2013}, EuNi$_2$P$_2$~\cite{Matsuda2009}, Sm$_{1-x}$Gd$_x$Al$_2$~\cite{Bersweiler2013} and SmB$_6$~\cite{Chen2018}. Moreover, through lineshape analysis of the Eu$^{2+}$ and Eu$^{3+}$ resonances  we observe that their resonant energy positions  exhibit a linear dependence as a function of pressure, as shown in Fig~2d. They show a pressure-induced compression effect as the their energy difference diminishes from $\sim$9.9~eV at lowest pressure of 1~GPa to $\sim$8.3~eV at the highest pressure of 48~GPa.  During this compression, the full width at half maximum(FWHM)  reveal the occurrence of an electronic phase transition. While the FWHM of Eu$^{2+}$ resonance remains unchanged for the whole pressure range, the FWHM of Eu$^{3+}$ resonance exhibit a sudden increase at 30~GPa, from about 3~eV to about 6~eV. This electronic phase transition leads naturally to a strong enhancement of spin-orbital interactions due to the activation of a large orbital momentum characteristic of the  Eu$^{3+}$ electronic state.\\

The normalized XMCD intensity of $A=A_{2+}+A_{3+}$ (Fig.~\ref{fig:3}b) shows a completely different behavior when compared to the mean valence value. It remains unchanged up to 10 GPa (region I) and is slightly decreasing (10\%) from 10 to 30~GPa (region II), followed by a sharp enhancement from 30 to 40~GPa (region III) with a factor of 3 and finally dropped from 42 to 48~GPa (region IV). The slightly reduced magnetization in region II, indicates for the continuously increase of the N{\'e}el temperature for moderate pressure after 8~Gpa~\cite{Uchima2014}. The jump of the macroscopic magnetization observed in region III clearly demonstrates the transition from AFM to a new magnetic phase at $\sim$30 GPa with a mean valence value of $\nu$=2.30. Following the increase of $\nu$ above 10 GPa, the magnetic contribution from Eu$^{3+}$ increases from 0.10 at $\sim$10 GPa to 0.20 at $\sim$48 GPa, as shown in Fig.~\ref{fig:3}c. This demonstrates that a stronger cumulative spin and orbital magnetic contribution from Eu$^{3+}$ state correlates with a higher Eu$^{3+}$ occupation in EuNiGe$_3$ under pressure. 

\section*{Discussion}\hspace*{\fill}\\
\\
The magnetic contributions from Ni sites are negligible as probed by in-situ high pressure XAS and XMCD  measured at the Ni K-edge up to 45.5 Gpa,  shown in Fig.~\ref{fig:sup6} and Fig.~\ref{fig:sup7} in the Supplementary materials. Besides, there is no structural phase transition  observed up to 57 Gpa, as demonstrated  by the in-situ high-pressure X-ray diffraction results shown in Fig.~\ref{fig:sup8} in the Supplementary materials. 

\begin{figure}[t]
\centering
\includegraphics[width=0.8\linewidth]{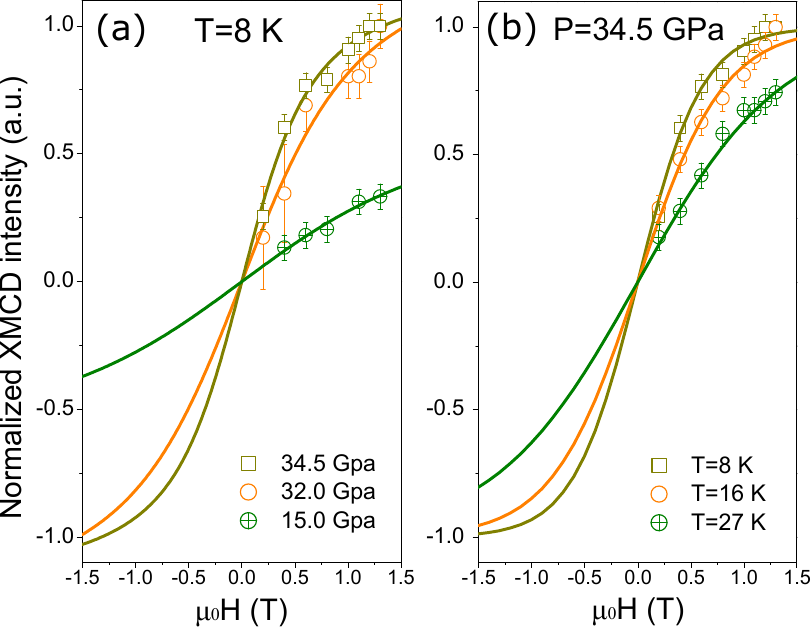}
\caption{\small{Field dependent XMCD intensity of Eu, obtained at T=8~K under pressure of 15.0, 32.0, and 34.5 GPa (a), and at T=8, 16, and 27~K under pressure of 34.5 GPa, the lines are superparamagnetic fittings with Brillouin function.} }    
 \label{fig:4}
\end{figure}

In addition to the large enhancement of the Eu magnetic polarization under pressure P$>$ 30.0~GPa, an onset of a specific change of magnetic behavior is observed according to the field dependence of the XMCD intensity of Eu$^{2+}$ at P=32.0 and 34.5~GPa, as shown in Fig.~\ref{fig:4}a. The profile of the XMCD spectra does not change with the field suggesting the same magnetic behavior of the 5d channels from Eu$^{2+}$ and Eu$^{3+}$. The saturation tendency and the S-shape magnetic hysteresis loop indicates the onset of a thermally activated dynamics of the magnetic state above 30 GPa. By contrast, an almost linear curve is observed for P=15.0 GPa in the AFM state, which is the ground state of the EuNiGe$_3$ at ambient pressure~\cite{Uchima2014}. For simplicity, we analyze this anomalous behavior in terms of a SPM model by fitting the field dependent XMCD with a Brillouin function~\cite{Jensen1991}. Note that the SPM model is most popular for the analysis of nanoparticles, whose magnetization can randomly flip direction within their characteristic relaxation times. However, short range correlated spins may exhibit similar characteristic dynamics in magnetic systems. In particular, dense spin- or magnetic cluster-glasses, spin-density-wave order under random exchange or random-field or other glassy magnetic systems do show such a behavior.\\

The first scenario includes ferromagnetic correlation that are available in the magnetic ground state. In our case, the ground state is antiferromagnet, therefore dipolar interactions and an eventual percolation threshold cannot be supported. For the second scenario, a transition from spin density waves to a glassy behavior would require a breaking symmetry mechanism which involves impurities and/or random exchange fields. This can also be excluded, because the EuNiGe3 is a single crystal (no impurities) and the ground state is antiferromagnetic (no random fields). Similar arguments applied also for the third scenario. Instead, as we mentioned above, the EuNiGe$_3$ crystal exhibits a non-centrosymmetric lattice structure, which allows for the presence of the antisymmetric DMIs~\cite{Dzyaloshinskii1958, Moriya1960}.Then it is reasonable that a pressure driven transition that involves valence fluctuations (Eu$^{2+}$/Eu$^{3+}$) under the presence of symmetry breaking interactions causes a transition from AFM to an unconventional superparamgnetic state.  This reflects short range interaction of spins that are characterized by an effective magnetic moment which fluctuates with a paramagnetic long range character.\\

Considering that the fluctuations are described by an effective moment, it defines the curvature of the field-dependent magnetization as a parameter. For 32.0~GPa an equivalent of 4 magnetic Eu$^{2+}$ state is reproducing the data whereas at 34.5~GPa an average number of 6.5 elemental moments result from the fitting to the data. These numbers, that reflect the ordered spins, are significantly higher as compared to a simple paramagnetic behavior where one magnetic atom would define the magnetization character. Corroborated also by the enhanced magnetization from 30.0 to 42.0~GPa (Fig.~\ref{fig:3}c) one can suggest that the short range magnetic interactions are strengthened by the lattice contraction, similar to that observed in EuX (X=Te, Se, S, O) monochalcogenides \cite{Souza-Neto2009} and $\rm Eu_{0.5}Yb_{0.5}Ga_4$ \cite{Loula2012}. For consistency check of the SPM behavior at high pressure we plot in Fig.~\ref{fig:4}b the XMCD dependence as a function of field for three different temperatures measured at 34.5~GPa. These curves also show the effect of enhanced thermally fluctuating moments by the change of curvature with the typical thermally activated SPM-like dynamics.\\

\begin{figure}[t]
\centering
\includegraphics[width=0.62\linewidth]{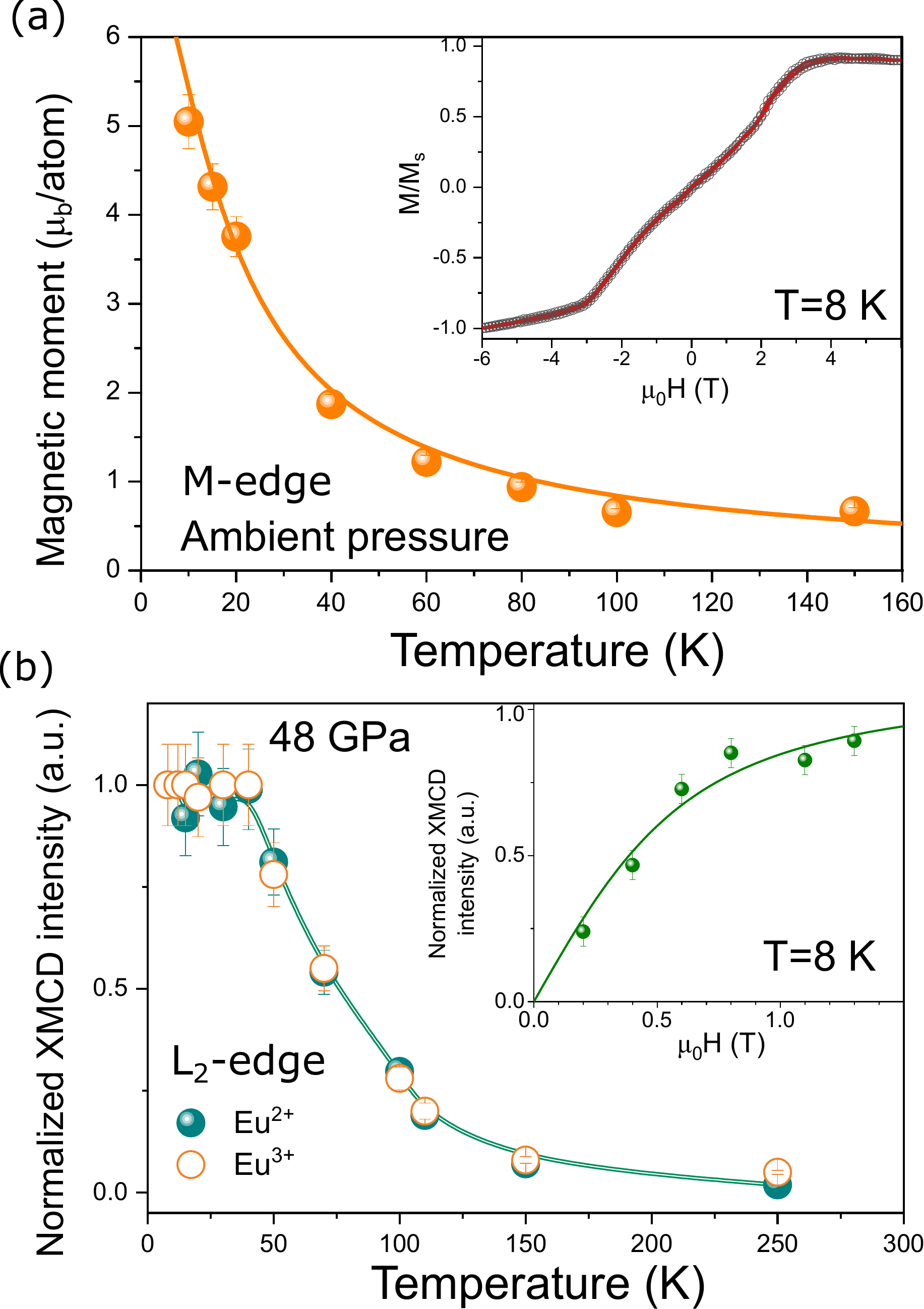}
\caption{\small{(a) Temperature dependence of the magnetic moment of Eu at ambient pressure. The line represent a plot of the Brillouin function for parameters characteristic to the divalent Eu. Inset: the AFM-type hysteresis loop measured at 8~K. (b) Eu$^{2+}$ and Eu$^{3+}$ XMCD intensities (normalized to the value of 8K) measured at  L$_2$-edge (solid and open circles) obtained at P=48.0~GPa from 8 to 250K and the line is a guide to the eyes. Inset: field dependent XMCD intensity of Eu$^{2+}$.} }    
 \label{fig:5}
\end{figure}

The paramagnetic behavior of EuNiGe$_3$ at ambient pressure and above the N{\'e}el temperature is confirmed according to the temperature dependent magnetic moments as shown in Fig.~\ref{fig:5}a.  For each temperature the XMCD has been measured at Eu M$_{4,5}$-edges in an external field of $\mu_0H=\pm 8~T$ applied along the c-axis of the crystal. The magnetic moments have been retrieved through the sum rules~\cite{cara} analysis applied to the XMCD spectra~(not shown). The line in Fig.~\ref{fig:5}a represents a plot of the Brillouin function for parameters characteristic to a divalent Eu. The agreement between the model and the measured magnetic moment as a function of temperature confirms the paramagnetic behavior of the magnetization above the ordering temperature. Below the N{\'e}el temperature, the hysteresis loop at T=8~K (inset of Fig.~\ref{fig:5}a) confirms its AFM ground state at low temperatures. In Fig.~\ref{fig:5}b we show the XMCD intensity (at L$_2$ edge) of Eu$^{2+}$ as well as Eu$^{3+}$ which were recorded under P=48.0 GPa, for an external field of $\mu_0H$=1.4~T and for temperatures ranging from 8 to 250K. The normalized values of $\rm A_{2+,3+}(T)/A_{2+,3+}(8K)$ deviate significantly from the ideal paramagnetic behavior, confirming a SPM character.  Also, the XMCD intensity of the Eu$^{3+}$ follows closely  the behavior of Eu$^{2+}$ which suggests a strong intra-atomic  exchange interaction in the valence-fluctuating EuNiGe$_3$.\\

The mechanism of the SPM-like state correlates with the onset of the electronic phase transition which leads to the onset of the spin-orbit coupling trough populating  the Eu$^{3+}$ electronic state. EuNiGe$_3$  has an acentric polar crystal structure (of BaNiSn$_3$ structure-type: space group I4mm, No.107) which causes the appearance of frustrating chiral Dzaloshinskii-Moriya interactions (DMIs) \cite{Dzyaloshinskii1964} that are enhanced by the onset of the orbital moment of the Eu$^{3+}$  at the transition pressure. This mechanism is present in EuNiGe$_3$ by symmetry and an unconventional  transition from AFM to another magnetic LRO or the paramagnetic state is expected to display an intermediate or meso-phase with fluctuating larger magnetic units than the paramagnetic ions. The chiral DMIs are always present, thus they are active also in the homogeneous intermediate valence state, where the on-site fluctuations between 4f$^7$ and 4f$^6$ configurations are so fast that magnetic properties are determined by the magnetic moments of a smeared state with fractional valence on site and its intersite exchange.\\

To conclude, element and orbital selective XAS and XMCD measurements on Eu L$_2$ absorption edges under pressures up to 48.0 GPa show a prominent valence change in EuNiGe$_3$ from Eu$^{2+}$ towards Eu$^{3+}$ as a function of pressure. Both the 5d channels of the Eu$^{2+}$ and Eu$^{3+}$ contributions are magnetically polarized and an electronic phase transition is observed at 30~GPa as a sudden increase of the resonance linewidth of the Eu$^{3+}$. Concomitantly, a magnetic transition to an anomalous state of slow and large thermal fluctuating moments is observed. The chiral magnetic exchange and a precursor state is identified as the underlying mechanism for this anomalous state. In EuNiGe$_3$, the 5d orbital channels of Eu$^{3+}$ has J=0 ground state and therefore is not responsive to the applied magnetic field. The polarization of 5d orbital channels of Eu$^{3+}$, which is intimately bound to that of Eu$^{2+}$ for all temperature and pressure ranges, suggests for intra-atomic exchange interactions to the Eu$^{2+}$ in valence-fluctuating EuNiGe$_3$. Such  strong intra-atomic exchange and spin-orbit interactions needs to be considered for future theoretical investigations of Eu- and other rare earth based materials with a valence fluctuating state.

%\begin{methods}\
\section*{Methods}
\hspace*{\fill}\\
\\
Single crystals of EuNiGe$_3$ were grown by using a high temperature solution growth method with In as a solvent, as described in more details in Refs.~\cite{Maurya2014,Kumar2010,Kumar2012}. The XAS and XMCD spectra at the Eu L$_2$-edge and Ni K-edge have been performed at ODE beamline~\cite{Baudelet2016} at synchrotron-SOLEIL, France to probe the pressure dependent local electronic configuration and 5d magnetism of Eu ions. Micrometer-sized powders ground from a high-quality single crystal EuNiGe$_3$, together with the pressure-transmitting medium silicon oil, was pressurized up to 48 GPa in a diamond-anvil cell. The pressure was measured using a ruby fluorescence scale. XMCD spectra were obtained through the difference of XAS spectra measured under the magnetic field up to $\mu_0$H=1.4~T, applied parallel or antiparallel to the beam helicity. The XMCD at the Eu M-edges were measured at VEKMAG end-station~\cite{RaduNoll2017} installed at the PM2 beamline of the synchrotron facility BESSY II, under external magnetic fields up to $\mu_0$H=8~T applied along c-axis of the single crystal. The in situ high-pressure X-ray diffraction measurement was performed with an angle-dispersive synchrotron X-ray diffraction mode (AD-XRD) at BL04 beamline of the ALBA.
%\end{methods}\\

\section*{Supporting Information}\hspace*{\fill}\\
\\
%\large\textbf{Supporting Information:}\normalsize 
Details of the preparation of EuNiGe$_3$ single crystal under investigation and its anisotropic magnetic properties in ambient pressure conditions, the element specific XMCD measurements for Eu and Ni, utilizing soft x-ray spectroscopy at the ambient pressure, the high-pressure XMCD at Ni K-edge, and the high-pressure X-ray diffraction results.
%\end{Supporting Information}

\section*{Acknowledgments}\hspace*{\fill}\\
\\
We acknowledge Synchrotron Soleil, HZB and ALBA for provision of synchrotron radiation facilities. Financial support for developing and building the PM2-VEKMAG beamline and VEKMAG end-station was provided by HZB and BMBF (Grants No. 05K10PC2, No. 05K10WR1,and No. 05K10KE1), respectively.F.R. acknowledge funding by the German Research Foundation via Project No. SPP2137/RA 3570. S. Rudorff is acknowledged for technical support.\\
\\
 The authors declare that they have no
competing financial interests.\\
 Correspondence should be addressed to K.~Chen(kaichen2021@ustc.edu.cn) and F.~Radu 
(florin.radu@helmholtz-berlin.de).
\section*{Author contribution}\hspace*{\fill}\\
\\
 K.C. and F.R. conceived, designed the projects, K.C., F.B. and F.R. performed the experiments. A.M. and A.T. prepared the single crystal samples. K.C., F.R., U.K.R.and D.M. co-wrote the paper. All the authors discussed the results and commented on the manuscript.

%% Here is the endmatter stuff: Supplementary Info, etc.
%% Use \item's to separate, default label is "Acknowledgements"

%\newpage

\clearpage
\section*{\Large Supplementary Information}\hspace*{\fill}\\
\section*{Supplementary Note 1: Crystal growth and sample characterization by x-ray diffraction}\hspace*{\fill}\\
\\
\renewcommand{\thefigure}{S\arabic{figure}}
\setcounter{figure}{0}
The details of the crystal growth and the anisotropic magnetic properties are discussed in Ref. [S1] of the main manuscript. Here we describe the preparation of EuNiGe3 single crystal under investigation and its anisotropic magnetic properties in ambient pressure conditions.  Also, we show for one temperature and at the ambient pressure the element specific XMCD measurements for Eu and Ni, utilizing soft x-ray spectroscopy. High-pressure X-ray diffraction results showing no phase transition up to 57.3 GPa.

\begin{figure}[h]
\centering
\includegraphics[width=.8\linewidth]{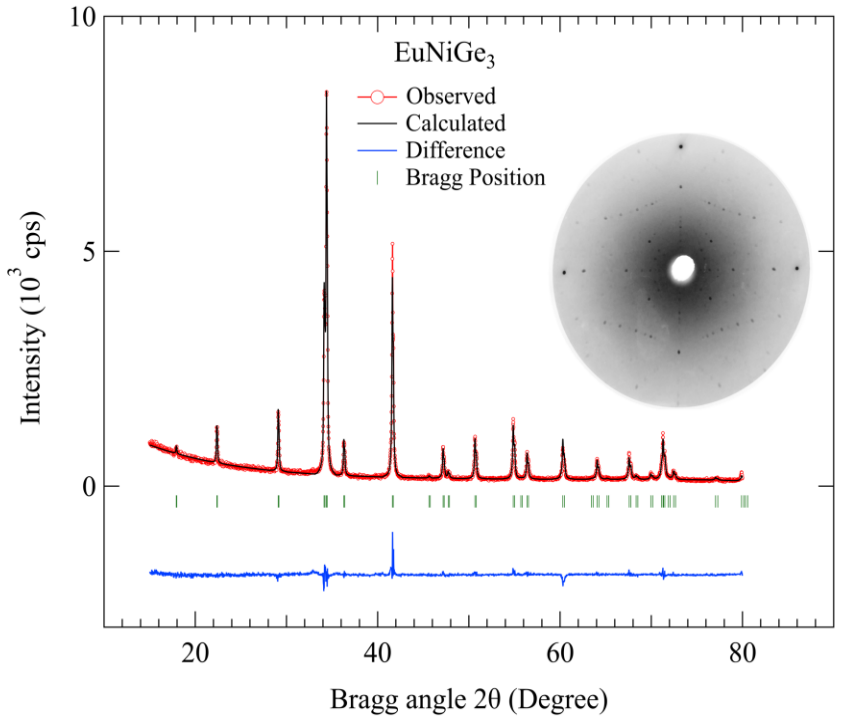}
\caption{\small{Field dependent XMCD intensity of Eu, obtained at T=8~K under pressure of 15.0, 32.0, and 34.5 GPa (a), and at T=8, 16, and 27~K under pressure of 34.5 GPa, the lines are superparamagnetic fittings with Brillouin function.} }    
 \label{fig:sup1}
\end{figure}

Single crystals of EuNiGe$_3$ were grown by high temperature solution growth using molten In as the flux. To start with, a polycrystalline sample of EuNiGe$_3$ was prepared by arc melting method and then this polycrystalline specimen along with excess indium (In) was placed in a recrystallized alumina crucible.  The crucible was subsequently sealed in a quartz ampoule and heated to a maximum temperature of 1100 $^\circ$C and held at this temperature for about 24 h, for proper homogenization.  Then the furnace was cooled down to 600 $^\circ$C at a rate of 2 $^\circ$C/h where the excess indium was centrifuged.  The quality of the crystals has been characterized by powder x-ray diffraction analysis, after powdering the some of the single crystals.  The x-ray diffraction pattern was clean without any impurity peaks, confirming the space group I4mm and the lattice constant.  Furthermore, the Laue diffraction was performed to orient the crystal along the crystallographic directions.  Well defined Laue diffraction pattern confirms the good quality of the single crystal.  The x-ray diffraction pattern, the Laue diffraction pattern, as well as the Rietveld refinement of the scattering pattern are shown in Fig.~\ref{fig:sup1}.

\section*{Supplementary Note 2:  Magnetic characterization of the  EuNiGe$_3$ at ambient pressure}
\begin{figure}[h]
\centering
\includegraphics[width=1\linewidth]{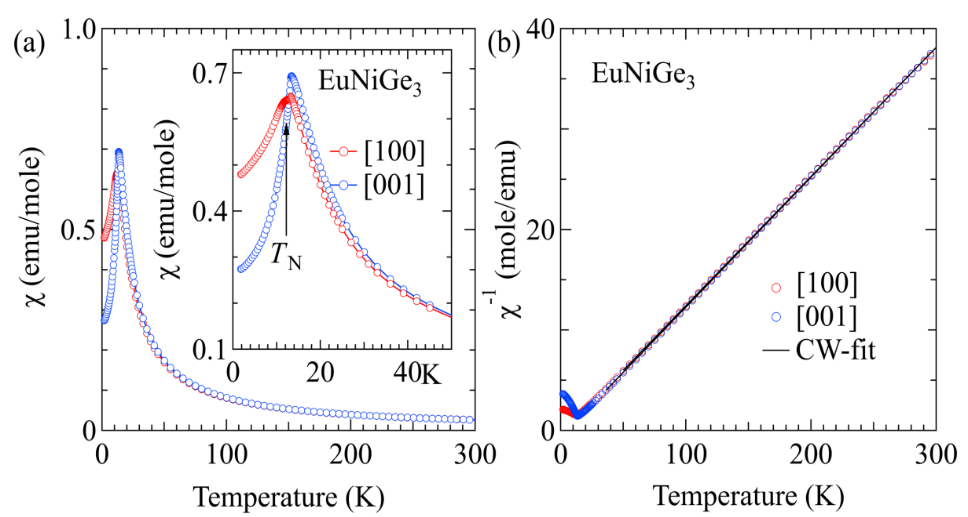}
\caption{\small{(a) Temperature dependence of magnetic susceptibility along the two principal crystallographic directions.  The inset shows the low temperature part, the Neel temperature is indicated by an arrow.  (b) Inverse magnetic susceptibility, the solid lines represent the Curie-Weiss fit. } }    
 \label{fig:sup2}
\end{figure}

\begin{figure}[ht]
\centering
\includegraphics[width=.8\linewidth]{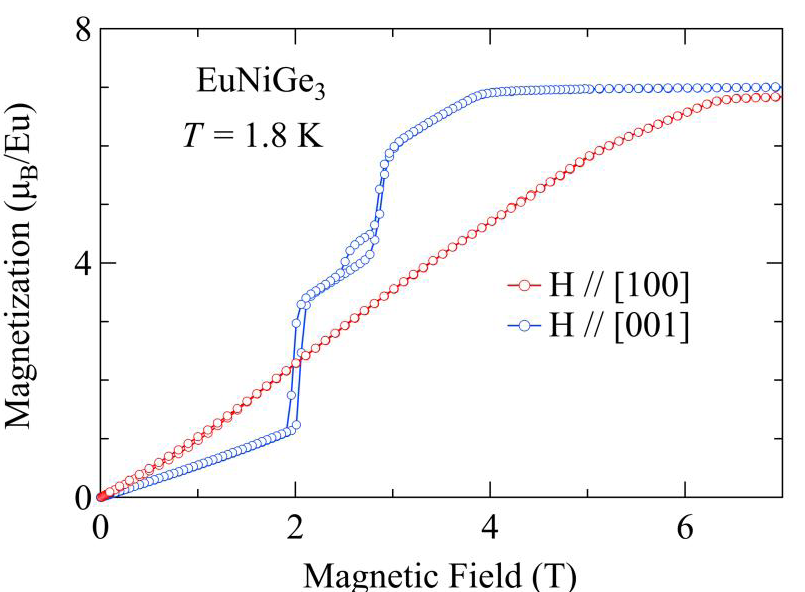}
\caption{\small{ 
Isothermal magnetization measured at T = 2 K along the two principal crystallographic directions.  The magnetization along H $\parallel$  [001] exhibits multiple metamagnetic transitions and saturates at around 4 T, while the magnetization along [100] direction is linear and saturates to 7 $\mu_B$/Eu at around 6 T. } }    
 \label{fig:sup3}
\end{figure}

\hspace*{\fill}\\
\\
The temperature dependent magnetic susceptibility from 2 to 300 K is shown in Fig.~\ref{fig:sup2}(a).  It is evident from the figure that in the paramagnetic region, the susceptibility follows the Curie-Weiss behavior, and a sharp cusp is observed at T$_N$ = 13.2 K.  The magnetic susceptibility drops rapidly for H $\parallel$ [001] direction confirming the easy axis of magnetization.   In the Fig.~\ref{fig:sup2}(b) we show the inverse magnetic susceptibility data.  A Curie-Weiss fit from 50 K to 300 K yields the paramagnetic Weiss temperature and the effective magnetic moment ($\mu_{eff}$) as 3.4 K and 7.89 $\mu_B$/Eu, respectively for H // [100] and 5.1 K and 7.90 $\mu_B$/Eu for H $\parallel$ [001] direction.  The effective moment value confirms that Eu exhibits a divalent state at ambient pressure conditions.  The isothermal magnetization measured at T = 2 K along the two principal crystallographic directions, is shown in Fig.~\ref{fig:sup3}.  For H $\parallel$ [100], the magnetization varies linearly with field up to 6.2 T, with a slight change of slope at around 4.6 T.  For H $\parallel$ [001] the magnetization increases linearly with field and undergoes a spin-flop transition at 2 T, followed by another spin re-orientation at 3 T, and finally another spin re-orientation at 4.1 T where it saturates to 7 $\mu_B$/Eu. 

\section*{Supplementary Note 3:  Eu and Ni L-edge XAS and XMCD spectra at ambient pressure and low temperature}\hspace*{\fill}\\
\\
Measurements of the single crystal have been performed by soft-x-ray absorption spectroscopy at the Eu M edges and Ni L edges. In order to access uncontaminated states of the probing elements, the sample was cleaved in-situ. This has been accomplished by fixing a 2mm in diameter Cu post on the crystal surface which has been detached in UHV conditions using a hard metal hit.  The temperature of the sample was set to ~12.5 K and the x-ray absorption spectra (XAS) were collected in external magnetic fields oriented perpendicular to the sample surface, along the c-axis.

\begin{figure}[ht]
\centering
\includegraphics[width=.8\linewidth]{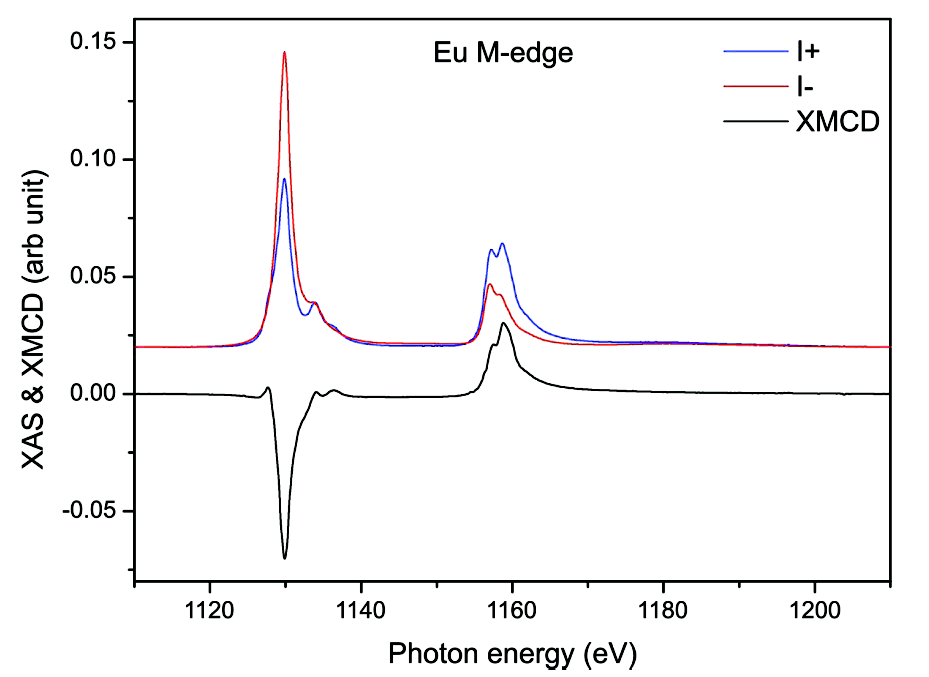}
\caption{\small{ XAS and XMCD spectra of Eu M-edge, measured at 10K, 8T for EuNiGe$_3$. } }    
 \label{fig:sup4}
\end{figure}

In Fig.~\ref{fig:sup4} we show the XAS spectra measured in Total electron yield mode for a magnetic field of $\mu_0$H=8T (blue line) and for an opposite magnetic field of $\mu_0$H=-8T. We observe a large difference of the spectra which provides, through their subtraction, the x-ray magnetic circular dichroism (XMCD) signal, which is shown as black line in the same figure. The lineshape of both XAS and XMCD are characteristic of a divalent Eu.

\begin{figure}[ht]
\centering
\includegraphics[width=.8\linewidth]{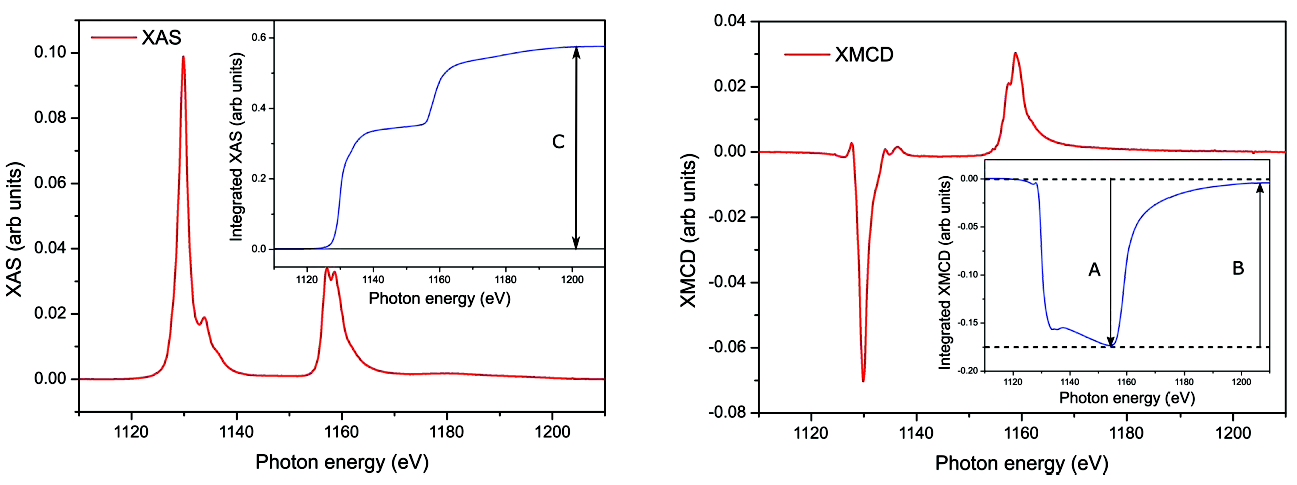}
\caption{\small{ Sum rule analysis of Eu M-edge, (left) XAS spectra and the integrated XAS spectra and (right) XMCD and the integrated XMCD spectra.} }    
 \label{fig:sup5}
 \end{figure}

By applying the sum rules (see Fig.~\ref{fig:sup5}) as described in Ref.[S3-S4] it is possible to determine the effective spin magnetic moment meffS  and the orbital magnetic moment m$_L$:
%\begin{equation}
$$M^{eff}_s =2<S_Z> + 7<T_Z>=-\frac{2A-3B}{2C} n_{4f}$$
$$M_L=-\frac{A+B}{C} n_{4f}
$$
%\end{equation}
where A and B are the integrated XMCD intensity at the Eu M5 and M4 edges, respectively (Fig.~\ref{fig:sup5} right panel). n4f is the number of holes in the 4f shell of Eu, C is the sum of the integrated XAS intensity at Eu M5 and Eu M4 edges (Fig.~\ref{fig:sup5} left panel), T$_Z$ is the magnetic-dipole operator of Eu. Assuming T$_Z$=0 for Eu2+ and considering n$_{4f}$=7 we obtained an almost zero orbital moment (ML ~ 0.01 $\mu_B$) and MS=5.2(0.3) $\mu_B$ per Eu2+ ion at 12.5 K. 

\begin{figure}[ht]
\centering
\includegraphics[width=.8\linewidth]{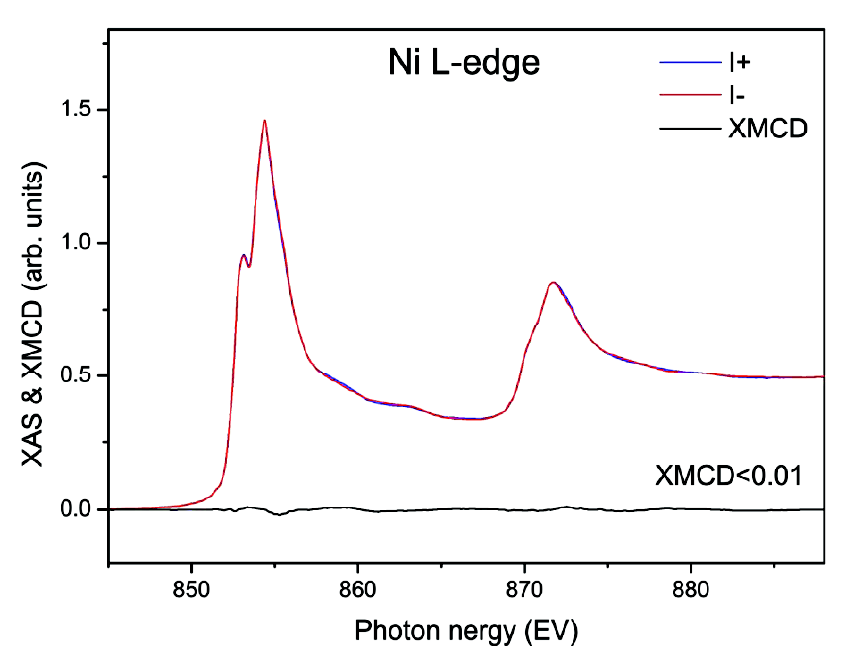}
\caption{\small{XAS and XMCD spectra of Ni L-edge, measure at 10 K, 8T for EuNiGe$_3$. The very small XMCD intensity (<1\%) suggests for the non-magnetic Ni sites in EuNiGe$_3$.} }    
 \label{fig:sup6}
\end{figure}

Ni L-edge XAS and XMCD spectra were recorded at 12.5 K, for magnetic fields of $\mu_0$H= $\pm$ 8T, applied along the c-axis (Fig.~\ref{fig:sup6}). The XAS curves are very similar to each other, suggesting for non-magnetic Ni atoms in EuNiGe$_3$, as proposed in Ref. [S2]. The XMCD signal, which is less than 1\% of the XAS intensity, is negligible as compared to the bulk Ni. The quench of the magnetization of Ni atoms confirms that the magnetic properties are ruled by the Eu atoms, alone in EuNiGe3.

\section*{Supplementary Note 4: Ni K-edge XAS and XMCD spectra at high pressure and low temperature}
\begin{figure}[ht]
\centering
\includegraphics[width=1\linewidth]{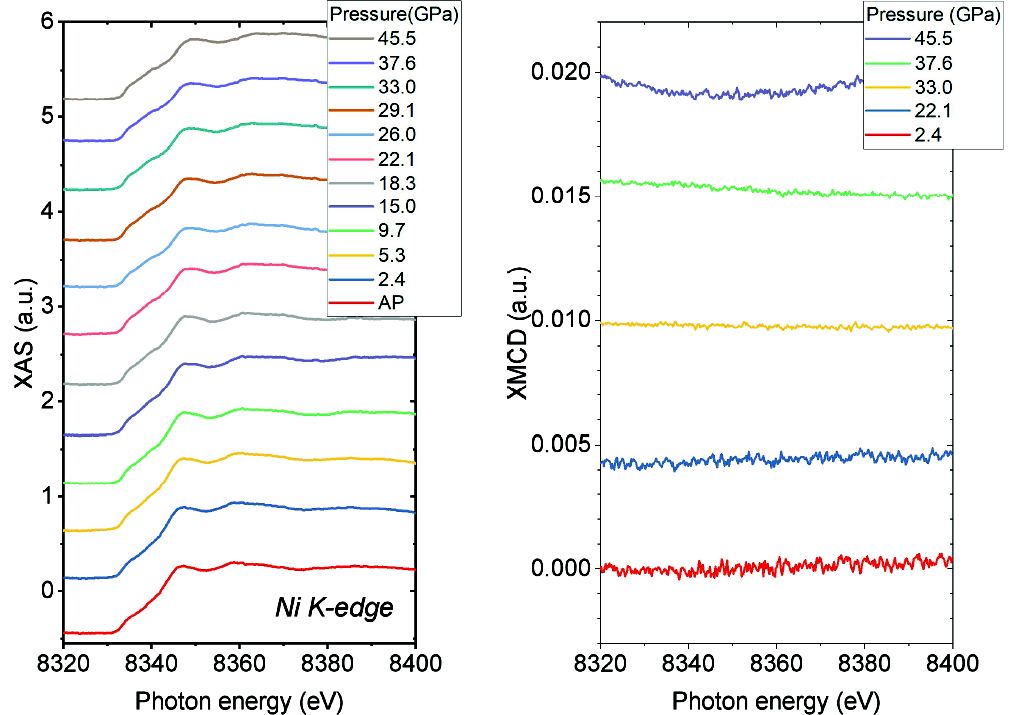}
\caption{\small{ In-situ high pressure XAS data of Ni K-edge up to 45.5 GPa (a) and the selected XMCD spectra(b) showing no magnetic contribution from Ni sites.} }    
 \label{fig:sup7}
\end{figure}

\hspace*{\fill}\\
\\
The in-situ high pressure XAS data of Ni K-edge up to 45.5 GPa and the selected XMCD spectra (See Fig.~\ref{fig:sup7}), which was performed at ODE beamline (H=1.3T and 10 K) at synchrotron Soleil, showing no magnetic contribution from Ni sites. Up on pressure increasing, the profile of the Ni K-edge XAS kept unchanged suggests for no structure transition of EuNiGe$_3$ materials.

\section*{Supplementary Note 5: High-Pressure X-ray Diffraction Results}
\begin{figure}[ht]
\centering
\includegraphics[width=1\linewidth]{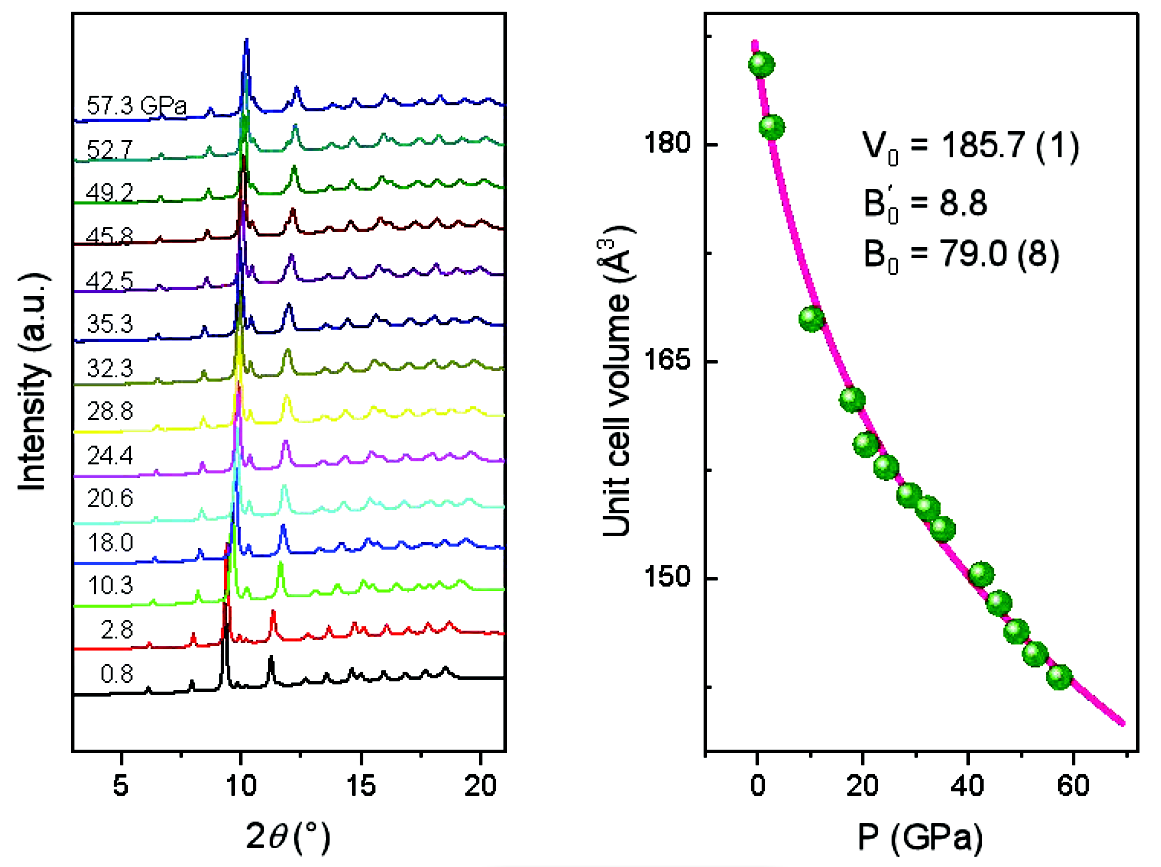}
\caption{\small{ Selected patterns of in-situ high pressure XRD data up to 57.3 GPa (a) and the V-P plots of EuNiGe$_3$ fitted with a third-order Birch-Murnaghan equation of state(b).} }    
 \label{fig:sup8}
\end{figure}

\hspace*{\fill}\\
\\
The {\textit{in situ}} high-pressure X-ray diffraction ($\lambda$= 0.4246 $\AA$) measurement was performed with an angle-dispersive synchrotron X-ray diffraction mode (AD-XRD) at beamline BL04 beamline of the ALBA. The as-prepared samples were loaded into a gasketed diamond anvil cell (DAC) with silicon oil as a pressure-transmitting medium for the highest pressure to 57.3 GPa. In-situ high pressure XRD measurements are performed and selected patterns are presented in Fig.~\ref{fig:sup8}(a), which did not reveal any crystallographic symmetry change up to 57.3 GPa. Nevertheless, one peak splitting near 12-degree occurred at 32.3 GPa, indicating a subtle structural deformation. The V-P plots of EuNiGe$_3$ was fitted with a third-order Birch-Murnaghan equation of state (EOS), as shown in Fig.~\ref{fig:sup8}(b), which yielded a bulk modulus of B$_0$ =79.0(8) GPa with B'= 8.8 up to the highest pressure. No phase transition is observed but accompanied with a lattice distortion above 32.3 GPa.\\
%\pagebreak
\section*{Supplementary References:}\hspace*{\fill}\\
\\
\lbrack S1\rbrack: Maurya, A.; Bonville, P.; Thamizhavel, A.; and Dhar, S. K.;  EuNiGe3, an anisotropic antiferromagnet, J. Phys.: Condens. Matter 2014, 26, 216001.\\
\lbrack S2\rbrack: Goetsch,  R. J.; Anand,  V. K.;  and Johnston,  D. C.; Antiferromagnetism in EuNiGe3, Phys. Rev. B 2013, 87, 064406. \\
\lbrack S3\rbrack: Thole, B. T.;  Carra, P.;  Sette, F.; and van der Laan, G.; X-ray circular dichroism as a probe of orbital magnetization, Phys. Rev. Lett. 1992, 68, 1943. \\
\lbrack S4\rbrack: Carra, P.; Thole, B. T.;  Altarelli, M.; and Wang, X.; X-ray circular dichroism and local magnetic fields, Phys. Rev. Lett. 1993, 70, 694.
\pagebreak

\end{document}